\documentclass[conference,10pt]{IEEEtran}
\usepackage{amsmath}
\usepackage{amssymb}
\usepackage{amsfonts}
\usepackage{graphicx}
\usepackage{enumerate}
\usepackage{bbm}
\usepackage{subfig}
\usepackage{mathtools} 
\usepackage{cite}
\usepackage{url}
\usepackage{array}
\usepackage{verbatim}
\usepackage{comment}
\usepackage{stfloats}
\usepackage[hyperindex]{hyperref}

\usepackage{hhline}

\usepackage[font=small,skip=0pt]{caption}
\DeclareMathOperator*{\argmax}{argmax}

\usepackage{color}
\definecolor{red}{rgb}{1,0,0}
\definecolor{blue}{rgb}{0,0,1}

\IEEEoverridecommandlockouts
\begin{document}
\title{Enhancing Physical Layer Security for NOMA Transmission in mmWave Drone Networks}
\author{
\IEEEauthorblockA{Nadisanka~Rupasinghe\IEEEauthorrefmark{1}, Yavuz~Yap{\i}c{\i}\IEEEauthorrefmark{1},
\. {I}smail~G\"uven\c{c}\IEEEauthorrefmark{1}, Huaiyu Dai\IEEEauthorrefmark{1}, Arupjyoti Bhuyan\IEEEauthorrefmark{2}
\IEEEauthorblockA{\IEEEauthorrefmark{1}Department of Electrical and Computer Engineering, North Carolina State University, Raleigh, NC}
\IEEEauthorblockA{\IEEEauthorrefmark{2}Idaho National Laboratory, Idaho Falls, ID}
{\tt \{rprupasi, yyapici, iguvenc, hdai\}@ncsu.edu,  arupjyoti.bhuyan@inl.gov}}%
\thanks{This research was supported in part by NSF under the grant CNS-1618692.}
}
\maketitle

\begin{abstract}

Physical layer security (PLS) is critically important for emerging wireless communication networks to maintain the confidentiality of the information of legitimate users. In this paper, we investigate enhancing PLS in an unmanned aerial vehicle (UAV) based communication network where a UAV acting as an aerial base station (BS) provides coverage in a densely packed user area (such as a stadium or a concert area). In particular, non-orthogonal multiple access (NOMA) together with highly-directional multi-antenna transmission techniques in mmWave frequency bands are utilized for improving spectral efficiency. In order to achieve PLS against potential eavesdropper attacks, we introduce a protected zone around the user region. However, limited resource availability refrain protected zone being extended to cover the entire eavesdropper region. Hence, we propose an approach to optimize the protected zone shape (for fixed area) at each UAV-BS hovering altitude. The associated secrecy performance is evaluated considering the secrecy outage and sum secrecy rates. Numerical results reveal the importance of protected zone shape optimization at each altitude to maximize NOMA secrecy rates.

\end{abstract}

\begin{IEEEkeywords}
5G, drone, HPPP, mmWave, non-orthogonal multiple access (NOMA), physical layer security (PLS), UAV.
\end{IEEEkeywords}

\section{Introduction}

The importance of deploying unmanned aerial vehicle (UAV) based communication networks during temporary events and after disasters to provide on-demand coverage and enhance capacity has recently been explored in some real word deployments and field trials~\cite{Att2,BBC}. In order to reap maximum benefits from such networks, enhancing the spectral efficiency (SE) is essential. To that end, integrating non-orthogonal multiple access (NOMA) transmission to UAVs acting as aerial base stations (BSs) can be an effective solution~\cite{NadisankaTCoM_arXiv, Nadisanka_SPAWC}. While enhancing the SE with NOMA, it is equally important to guarantee the confidentiality of communication going on between UAV-BS and legitimate users. Hence, introducing appropriate physical layer security (PLS) techniques to such networks become paramount importance.    



A UAV based mobile cloud computing system is proposed in \cite{7932157} where UAVs offer computation offloading opportunities to mobile stations (MS) with limited local processing capabilities. In that, just for offloading purposes between a UAV and the MSs, NOMA is proposed as one viable solution. In our earlier work \cite{Nadisanka_SPAWC, NadisankaTCoM_arXiv}, NOMA transmission is introduced to UAVs acting as aerial BSs to provide coverage over a stadium or a concert scenario. In particular, leveraging multi-antenna techniques a UAV-BS generates directional beams, and multiple users are served within the same beam employing NOMA transmission. In \cite{NadisankaTCoM_arXiv}, assuming the availability of user distance information, a beam scanning strategy is proposed to maximize NOMA sum rates whereas in \cite{Nadisanka_SPAWC} we study the performance of different feedback schemes for NOMA.

PLS in wireless communication networks has recently attracted significant attention \cite{PhySecrecy13_Zhang, PhySecurity13_Zurita,NOMAPhySecurity17_Yuanwei}. One of the main objectives of the PLS is to increase the performance gap of the link quality between the legitimate user and that of the eavesdropper (Eve) by exploiting the physical properties of the wireless medium\cite{PhySecuritySurvey18_Hamamreh}. To enhance the PLS in wireless ad-hoc networks, artificial noise (AN) aided multi-antenna transmission strategy is proposed in \cite{PhySecrecy13_Zhang}.
In~\cite{PhySecurity13_Zurita}, a \emph{protected zone} is defined surrounding the transmitter along with beamforming and AN transmission to enhance the PLS in a multi-input-single-output (MISO) communication system. Within the protected zone it is guaranteed that no Eve exists.  Considering single antenna and multi-antenna scenarios, PLS with NOMA transmission in large-scale networks is investigated in \cite{NOMAPhySecurity17_Yuanwei}. In particular, for single antenna scenario Eve exclusion area is proposed while for multi-antenna scenario AN generation towards undesired directions is introduced to enhance PLS.

In this paper, we consider a similar scenario as in \cite{Nadisanka_SPAWC,NadisankaTCoM_arXiv}, where a UAV-BS is employed to provide broadband connectivity over a densely packed user area in a stadium. NOMA along with multi-antenna transmission is then introduced to improve the SE. In particular, we consider there are Eves outside of the user area trying to breach communication going on between legitimate users and UAV-BS. In order to enhance the PLS of the UAV based communication network, we introduce a protected zone around the user area \cite{PhySecurity13_Zurita, NOMAPhySecurity17_Yuanwei}. However, due to physical constraints, protected zone may not be able to eliminate all the Eves distributed within the area. Hence, we propose an approach to optimize the protected zone shape based on UAV-BS hovering altitude such that the achievable NOMA sum secrecy rates are maximized.

\section{System Model} \label{sec:Sys_Model}
\subsection{Overview}

\begin{figure}[!t]
\begin{center}
\includegraphics[width=0.45\textwidth]{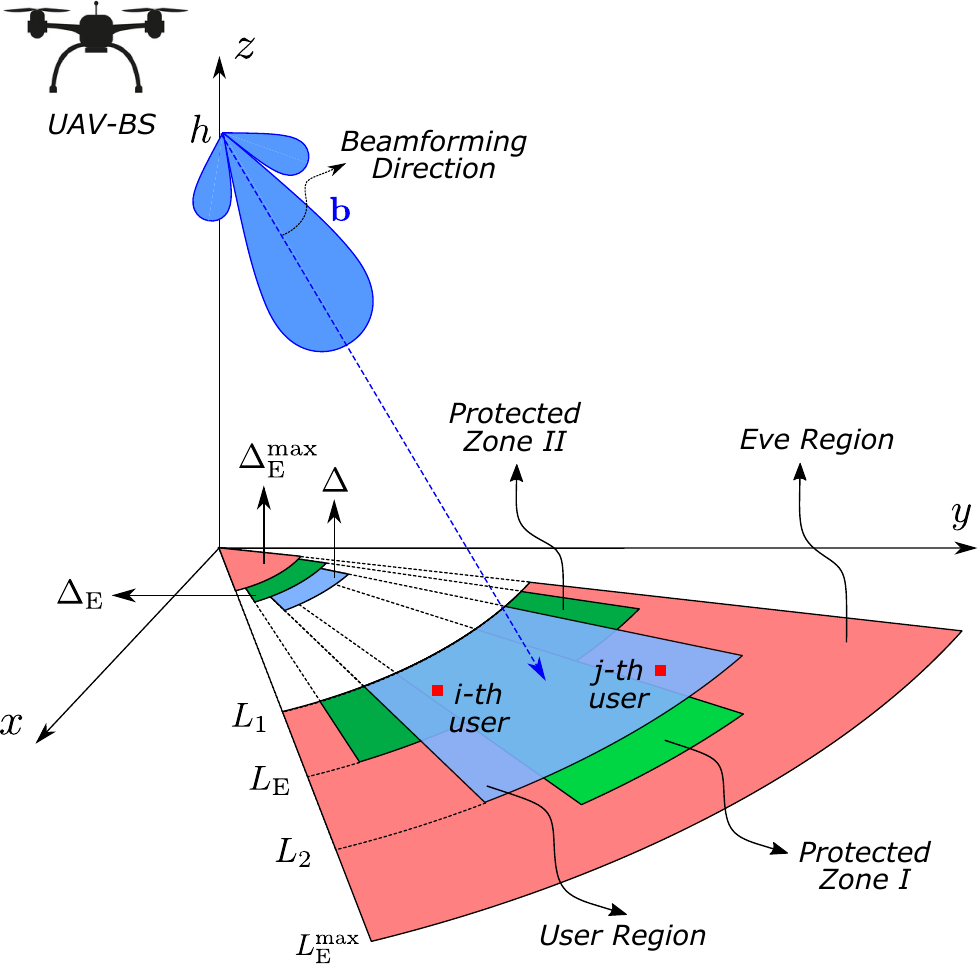}
\end{center}
\caption{System scenario where NOMA transmission serves multiple users simultaneously in a single DL beam.}
\label{fig:footprint}
\end{figure}

We consider a mmWave-NOMA transmission scenario where a single UAV-BS equipped with an $M$~element uniform linear array (ULA) is serving single-antenna users in the DL. We assume that all the users lie inside a specific \emph{user region} as shown in Fig.~\ref{fig:footprint}. A 3-dimensional (3D) beam is generated by the UAV-BS which entirely covers the user region. We assume that there are $K$ users in total, and the users can be represented by the set $\mathcal{N}_{\rm U} = \{1,2,\ldots K\}$. The user region is identified by an inner-radius $L_1$, an outer-radius $L_2$, and the angle $\Delta$, which is the fixed angle within the projection of horizontal propagation pattern of UAV-BS on the $xy$-plane. Note that it is possible to reasonably model various different hot spot scenarios such as a stadium, concert hall, traffic jam, and urban canyon by modifying these control parameters.

We assume that although the user region is free from eavesdroppers, the surrounding region includes Eves trying to intercept the transmission between UAV-BS and the legitimate users. We designate the bounded region around the user region, which includes Eves as \textit{Eve region}. Similar to the user region, we identify the Eve region by the same inner radius $L_1$, an outer radius $L_{\rm E}^{\rm max}$ (greater than $L_2$), and $\Delta_{\rm E}^{\rm max}$ (greater than $\Delta)$, as shown in Fig.~\ref{fig:footprint}. We assume $K_{\rm E}$ Eves in total, which are represented by the set $\mathcal{N}_{\rm E} = \{1,2,\ldots K_{\rm E}\}$. Note that horizontal footprint of the UAV-BS beam pattern covers the Eve region (so that any Eve has nonzero channel to UAV-BS), as well, but the coverage over Eve region might be provided by the side lobes depending on the specific radiation pattern.

\subsection{Location Distribution and mmWave Channel Model}
We assume that users and Eves are uniformly, randomly distributed within their specified regions following homogeneous Poisson point process (HPPP) with the densities $\lambda$ and $\lambda_{\rm E}$, respectively. The number of users (Eves) in the user (Eve) region is therefore Poisson distributed, i.e., $\textrm{P}(k \textrm{ users in the user region})\,{=}\, \frac{\mu^k e^{{-}\mu}}{k!} $ with $\mu\,{=}\,(L_2^2\,{-}\,L_1^2)\frac{\Delta}{2} \lambda$.

We assume that all the users have line-of-sight (LoS) paths since i) UAV-BS is hovering at relatively high altitudes, and ii) LoS path is much stronger than non-LoS (NLoS) paths in mmWave frequency band~\cite{Ding17PoorRandBeamforming,NadisankaTCoM_arXiv}. The channel $\textbf{h}_k$ between the $k$-th user and the UAV-BS is therefore given as
\begin{align} \label{eq:k_UE_original_channel}
\textbf{h}_k = \sqrt{M} \frac{\alpha_k \textbf{a}(\theta_k)}{\left[\textrm{PL}\left(\sqrt{d_k^2 + h^2}\right)\right]^{1/2}},
\end{align}
where $h$, $d_k$, $\alpha_{k}$ and $\theta_{k}$ represent UAV-BS hovering altitude, horizontal distance between $k$-th user and UAV-BS, small scale fading gain (i.e., complex Gaussian with $\mathcal{CN}(0,1)$), and angle-of-departure (AoD), respectively. In addition, $\textbf{a}(\theta_{k})$ is the steering vector associated with AoD $\theta_{k}$, and $\textrm{PL}\left(x\right)$ represents the path loss (PL) over the distance $x$. Note that the channel between $\ell$-th Eve in the Eve region (i.e., $\ell\,{\in}\,\mathcal{N}_{\rm E}$) and UAV-BS can also be given using \eqref{eq:k_UE_original_channel}.

\subsection{Protected Zone Approach for Physical Layer Security}\label{Sec:Protected_Zone}

\begin{figure}[!t]
\begin{center}
\includegraphics[width=0.4\textwidth]{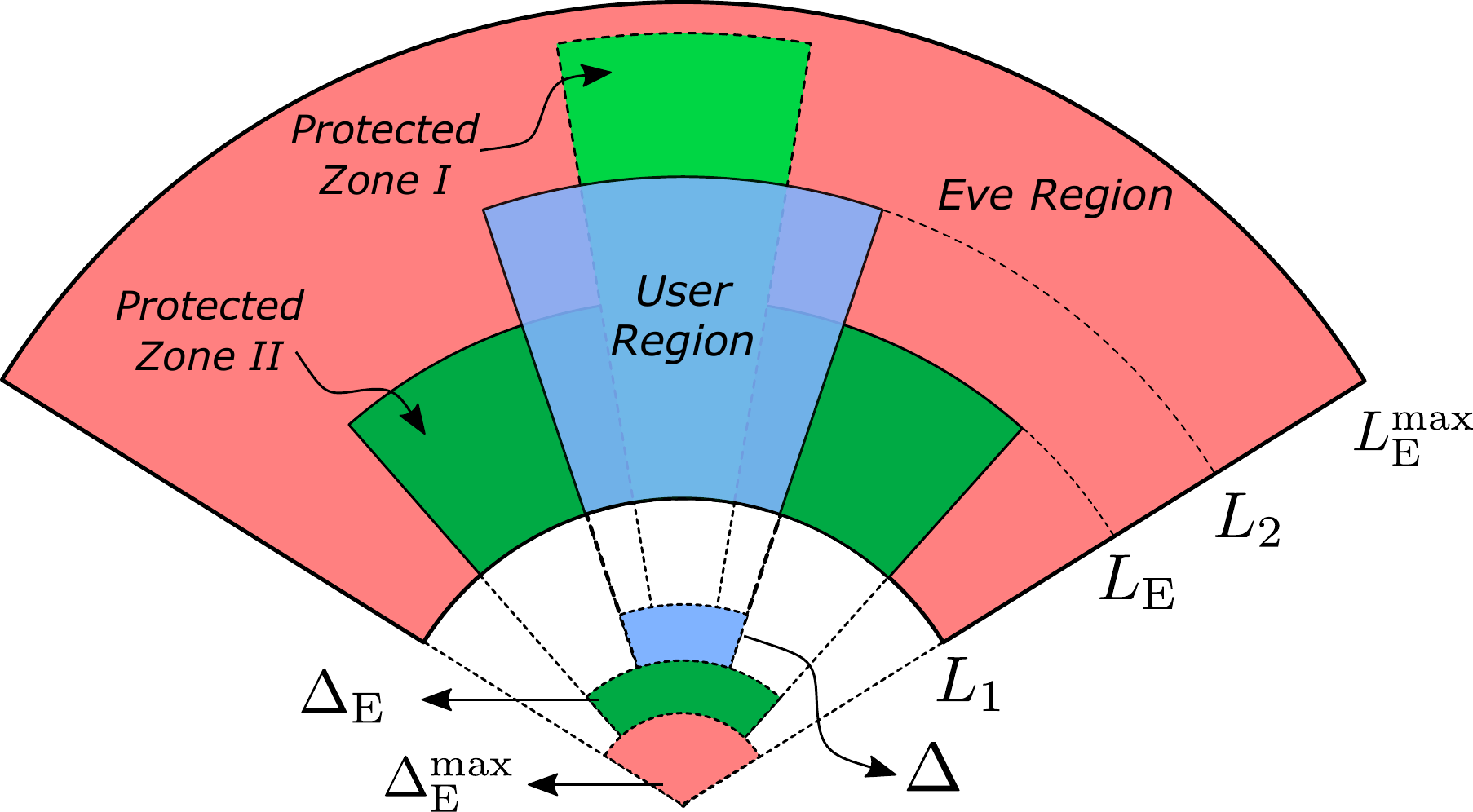}
\end{center}
\caption{Footprint of \emph{protected zone} represented by angle-distance pair $(\Delta_{\rm E},\, L_{\rm E})$, which is free from any eavesdroppers.}
\label{fig:ProtectedZone}
\end{figure}

The overall transmission scheme between the UAV-BS and legitimate users presented in Fig.~\ref{fig:footprint} is highly prone to the Eve attacks, and the PLS is accordingly impaired. In this study, we consider protected zone approach to enhance the secrecy rates of the network~\cite{PhySecurity13_Zurita, NOMAPhySecurity17_Yuanwei}. In the proposed approach, an additional area (i.e., protected zone) around the user region (and inside the Eve region) has been cleared from Eves by means of some measures, as shown Fig.~\ref{fig:ProtectedZone}. This protected area is actually a fraction of the complete Eve region, and we denote this fraction by $q$ with $q \,{\leq}\, 1$. Note that since clearing Eves in the protected zone requires certain resources being spent on the ground, our goal is to keep this area as small as possible. In addition, we consider to optimize the shape of the protected zone to enhance secrecy rates while keeping its area the same, which is the main problem we tackle in this study.

The protected zone can be represented by an angle-distance (radius) pair $(\Delta_{\rm E},\, L_{\rm E})$ with $\Delta_{\rm E}^{\min} \,{\leq}\, \Delta_{\rm E} \,{\leq}\, \Delta_{\rm E}^{\max}$ and $L_1 \,{\leq}\, L_{\rm E} \,{\leq}\, L_{\rm E}^{\max}$. Note that  $\Delta_{\rm E}^{\min}$ is the minimum angle value which occurs when $L_{\rm E} \,{=}\, L_{\rm E}^{\max}$. We can therefore represent $\Delta_{\rm E}^{\min}$ as follows
\begin{align} \label{eq:Angle_min_Delta}
\Delta_{\rm E}^{\min}=\frac{q\Big[\big((L_{\rm E}^{\max})^2\,{-}\,L_1^2\big)\Delta_{\rm E}^{\max}\,{-}\,(L_2^2\,{-}\,L_1^2)\Delta  \Big]}{(L_{\rm E}^{\max})^2\,{-}\,L_2^2}.
\end{align}

As sketched in Fig.~\ref{fig:ProtectedZone}, it is possible to have different shapes for protected zone for a fixed $q$ value. Note that whenever we have $\Delta_{\rm E} \,{\leq}\, \Delta$, $L_{\rm E}$ should be sufficiently greater than $L_2$ (e.g., ``Protected Zone I'' in Fig.~\ref{fig:ProtectedZone}) to have a nonzero protected zone. When $\Delta \,{\leq}\, \Delta_{\rm E} \,{\leq}\, \Delta_{\rm E}^{\max}$, $L_{\rm E}$ might however be smaller (e.g., ``Protected Zone II'' in Fig.~\ref{fig:ProtectedZone}) or greater than $L_2$ depending on the area of the user region and particular $q$ choice. Specifically, $L_{\rm E}$ can be parametrically expressed as follows
\begin{align} \label{eq:Le_less_Delta}
L_{\rm E}^2\,{=}\,L_2^2{+}\frac{q}{\Delta_{\rm E}}\Big[\big((L_{\rm E}^{\max})^2\,{-}\,L_1^2\big)\Delta_{\rm E}^{\max}\,{-}\,(L_2^2\,{-}\,L_1^2)\Delta  \Big],
\end{align}
for $\Delta_{\rm E}^{\min} \,{\leq}\, \Delta_{\rm E} \,{\leq}\, \Delta$. Whenever we have $\Delta \,{<}\, \Delta_{\rm E} \,{\leq}\, \Delta_{\rm E}^{\max}$,
\begin{align} \label{eq:Le_greater_Delta_1}
L_{\rm E}^2{=}L_1^2{+}\frac{q}{\Delta_{\rm E}}\Big[\big((L_{\rm E}^{\max})^2{-}L_1^2\big)\Delta_{\rm E}^{\max}{+}\frac{1{-}q}{q}(L_2^2{-}L_1^2)\Delta \Big],
\end{align}
provided $L_{\rm E}^2 \,{\geq}\, L_2^2$, and $L_{\rm E}$ is otherwise expressed as
\begin{align} \label{eq:Le_greater_Delta_2}
L_{\rm E}^2\,{=}\,L_1^2{+}\frac{q}{\Delta_{\rm E}{-}\Delta}\Big[\big((L_{\rm E}^{\max})^2{-}L_1^2\big)\Delta_{\rm E}^{\max}{-}(L_2^2{-}L_1^2)\Delta  \Big].
\end{align}

\section{Secure NOMA for UAV-BS Downlink} \label{Sec:NOMA_Transmission}

In this section, we consider NOMA transmission in UAV-BS downlink (DL) to enhance the SE, and evaluate the associated secrecy rates in the presence of protected zone.

\subsection{Secrecy Outage and Sum Secrecy Rates} \label{sec:Outage_Prob_Sum_Rates}

We assume that UAV-BS generates a beam $\textbf{b}$ where the respective projection in the azimuth domain is in the direction of $\overline{\theta}$ with $\overline{\theta} \,{\in}\, [0{,}\,2\pi]$ \cite{NadisankaTCoM_arXiv}.
Assuming critically spaced array, the effective channel gain of user $k\,{\in}\,\mathcal{N}_\textrm{U}$ for beamforming direction $\overline{\theta}$ can be given using \eqref{eq:k_UE_original_channel} as follows~\cite{Nadisanka_SPAWC, NadisankaTCoM_arXiv}
\begin{align} \label{eq:Eff_channel_gain}
|\textbf{h}_k^{\rm H}\textbf{b}|^2 &\approx \frac{|\alpha_k|^2}{M \times\textrm{PL}\left(\sqrt{d_k^2 + h^2}\right)}
 \left| \frac{ \sin \left( \frac{\pi M(\overline{\theta} - \theta_k)}{2} \right)}{  \sin \left( \frac{\pi (\overline{\theta} - \theta_k)}{2} \right)}\right|^2 ,
\nonumber \\
 &= \frac{|\alpha_k|^2}{\textrm{PL}\left(\sqrt{d_k^2 + h^2}\right)} {\rm F}_M(\pi [\overline{\theta} - \theta_k]),
\end{align} where ${\rm F}_M(\cdot)$ is the Fej\'er kernel. Similarly, the effective channel gain of the most detrimental Eve, $g_{\rm E}$ is given as, \begin{align} \label{eq:Eff_channel_gain_Eve}
g_{\rm E} = \max _{k_{\rm E}\in \mathcal{N}_{\rm E}} |\textbf{h}_{k_{\rm E}}^{\rm H}\textbf{b}|^2
\end{align}
where $\textbf{h}_{k_{\rm E}}$ is the channel gain of the $k_{\rm E}$-th Eve.

When deriving secrecy rates in NOMA transmission, we assume that UAV-BS knows the effective channel gains of desired users while those of Eves are unknown. Without any loss of generality, we also assume that the users in set $\mathcal{N}_{\rm U}$ are already indexed from the best to the worst with respect to their effective channel gains as represented by \eqref{eq:Eff_channel_gain}. Defining $\beta_k$ to be the power allocation coefficient of $k$-th user, we therefore have $\beta_1\,{\leq}\,\dots\,{\leq}\,\beta_K$ such that $\sum _{k{=}1}^{K} \beta_k^2\,{=}\,1$. The transmitted signal is generated by superposition coding as
\begin{align}
\textbf{x} = \sqrt{P_{\rm Tx}}\textbf{b}\sum \limits_{k = 1}^{K} \beta_k s_{k}, \end{align} where $P_{\rm Tx}$ and $s_{k}$ are the total DL transmit power and $k$-th user's message, respectively. The received signal at the $k$-th user is then given as
\begin{align} \label{eq:k-th_user_Rx_signal}
y_{k}= \textbf{h}_{k}^{\rm H} \textbf{x} +  v_k = \sqrt{P_{\rm Tx}}\textbf{h}_{k}^{\rm H} \textbf{b}\sum \limits_{k = 1}^{K} \beta_k s_{k} + v_k,
\end{align}
where $v_k$ is zero-mean complex Gaussian additive white noise with variance $N_0$.

With the received signal as in \eqref{eq:k-th_user_Rx_signal} in hand, each user first decodes messages of all weaker users (allocated with larger power) sequentially in the presence of stronger users' messages (allocated with smaller power). Those decoded messages are then subtracted from the received signal in \eqref{eq:k-th_user_Rx_signal}, and each user decodes its own message treating the stronger users' messages as noise. This overall decoding process is known as successive interference cancellation (SIC), and $k$-th user decodes its own message after SIC with the following SINR:
\begin{align} \label{eq:SINR_k_th_user}
{\rm SINR}_{k} =  \frac{P_{\rm Tx}|\textbf{h}_{k}^{\rm H}\textbf{b}|^2 \beta_{k}^2}{ \left( 1- \delta_{k1} \right) P_{\rm Tx}  \sum \limits_{l = 1}^{k-1}|\textbf{h}_{k}^{\rm H}\textbf{b}|^2 \beta_{l}^2 + N_0},
\end{align}
where $\delta_{k1}$ is the Kronecker delta function taking $1$ if $k\,{=}\,1$, and $0$ otherwise. Assuming that Eves have powerful detection capability \cite{NOMAPhySecurity17_Yuanwei, PhySecrecy13_Zhang}, the most detrimental Eve decodes $k$-th user message with the SINR given as
\begin{align} \label{eq:SINR_Eve_k_th_user}
{\rm SINR}_{k}^{\rm E} =  \frac{P_{\rm Tx}\beta_{k}^2 g_{\rm E} }{ \left( 1- \delta_{k1} \right) P_{\rm Tx}  \sum \limits_{l = 1}^{k-1} \beta_{l}^2 g_{\rm E} + N_0^{\rm E}},
\end{align}
where $N_0^{\rm E}$ is the associated noise variance.


Considering SINR in \eqref{eq:SINR_k_th_user}, the instantaneous rate at $k$-th user is given by $R^{\textrm{NOMA}}_k\,{=}\,\log_2(1+{\rm SINR}_{k})$. Similarly, considering \eqref{eq:SINR_Eve_k_th_user}, the instantaneous rate at the most detrimental Eve for decoding the $k$-th user message is given as $R^{\textrm{NOMA}}_{k,{\rm E}}\,{=}\,\log_2(1+{\rm SINR}_{k}^{\rm E})$. The secrecy rate for $k$-th legitimate user can therefore be given as \cite{PhySecurity08_Bloch, NOMAPhySecurity17_Yuanwei}
\begin{align} \label{eq:Secrecy_rate_k_th_user}
C^{\textrm{NOMA}}_k = \left[R^{\textrm{NOMA}}_k- R^{\textrm{NOMA}}_{k,{\rm E}}\right]^{+},
\end{align}
where $[x]^{+}\,{=}\,\max\{x,0\}$. As \eqref{eq:Secrecy_rate_k_th_user} implies, the secrecy rates are always strictly positive \cite{PhySecurityAntCorr13_Yang}.


Assuming that $\overline{R}_k$ denotes desired secrecy rate for the user $k \,{\in}\, \mathcal{N}_{\rm U}$, we define the secrecy outage event occurring whenever $C^{\textrm{NOMA}}_k \,{<}\, \overline{R}_k$ with the respective secrecy outage probability $\textrm{P}_{k}^{\rm o} \,{=}\, \textrm{P}\{C^{\textrm{NOMA}}_k \,{<}\, \overline{R}_k\}$. As a result, outage sum secrecy rate with NOMA transmission can be given as
\begin{align} \label{eq:sum_secrecy_rate_NOMA}
R^{\textrm{NOMA}} = \sum \limits_{k = 1}^{K}  (1- \textrm{P}_{k}^{o}) \overline{R}_k.
\end{align}
For performance comparison, we also consider outage sum secrecy rate with OMA transmission.

\subsection{Shape Optimization for Protected Zone} \label{sec:Shape_Optimization}

In this section, we discuss optimization of the protected zone shape to enhance the secrecy rates while keeping its area (i.e., $q$) the same. We note that any particular subregion within the Eve region does not equally impair the achievable secrecy rates even if the subregion areas are the same and the Eves are equally capable. This is basically due to the varying effective channel gain between UAV-BS and Eve with different subregions, which is a function of not only the distance but also the \textit{relative angle} (i.e., angle offset from the beamforming direction) associated with each Eve.

Considering \eqref{eq:Secrecy_rate_k_th_user}, the subregion involving the most detrimental Eve has the largest impact on the secrecy rates.
Hence, instead of choosing the subregions arbitrarily to form the protected zone, it is more meaningful to include (i.e., \textit{protect}) subregions which result in better effective channel gain for potential eavesdroppers, and hence is likely to involve the most detrimental Eve.

As we will show in Section~\ref{sec:Numerical-Location_Dist}, the location distribution of the most detrimental Eve depends also on the hovering altitude of UAV-BS. In particular, the most detrimental Eve is likely to be present in a subregion where $\Delta_{\rm E} \,{\geq}\, \Delta$ and $L_{\rm E} \,{\leq}\, L_2$, which is represented by ``Protected Zone II'' in Fig.~\ref{fig:ProtectedZone}, when the altitude is low. In contrast, the region including the most detrimental Eve becomes closer to ``Protected Zone I'' of Fig.~\ref{fig:ProtectedZone} with $\Delta_{\rm E} \,{\leq}\, \Delta$ and $L_{\rm E} \,{\geq}\, L_2$ when the altitude is high. We therefore conclude that the shape of the protected zone should be optimized taking into account the UAV-BS hovering altitude. Hence, at a particular altitude and for a given $q$, the optimal shape of the protected zone can be identified as
\begin{align}
\Delta_{\rm E}^*,\,L_{\rm E}^* = & \argmax_{\Delta_{\rm E},\, L_{\rm E}}\, R^{\textrm{NOMA}} \\
& \text{s.t.} \; \Delta_{\rm E}^{\min} \leq \Delta_{\rm E} \leq \Delta_{\rm E}^{\max}, \nonumber\\
& \hphantom{s.t.} L_{\rm E} \text{ is computed by } \eqref{eq:Le_less_Delta}{-} \eqref{eq:Le_greater_Delta_2},\nonumber
\end{align}
where $R^{\textrm{NOMA}}$ is given in \eqref{eq:sum_secrecy_rate_NOMA}. 

\section{Numerical Results} \label{sec:Numerical_results}

In this section, we present numerical results to show the importance of shape optimization of the protected zone and its impact on the achievable sum secrecy rates with varying UAV-BS hovering altitudes. Considering Fig.~\ref{fig:footprint}, we assume that $L_{2}\,{=}\,100$~m, $L_{1}\,{=}\,25$~m, $L_{\rm E}^{\textrm{max}}\,{=}\,1.5L_{2}$~m, $\Delta\,{=}\,0.02 \ \textrm {rad} \left( 1.145^{\circ}\right)$, $\Delta_{\rm E}^{\textrm{max}}\,{=}\,2\Delta$,  $\bar{\theta}\,{=}\,0^{\circ}$, and $M\,{=}\,100$. User distribution is based on HPPP with $\lambda\,{=}\,1$, and user target secrecy rates are $\overline{R}_j\,{=}\,4$~bits per channel use (BPCU) and $\overline{R}_i\,{=}\,1$~BPCU, respectively. The power allocation ratios are $\beta_j^2\,{=}\,0.25$ and $\beta_i^2\,{=}\,0.75$ while $P_{\rm Tx}\,{=}\,10$~dBm and $N_0\,{=}\,-35$~dBm. We assume two user NOMA transmission with $j\,{=}\ 1$ and $i\,{=}\ 20$ after ordering users with respect to their effective channel gains. The path-loss model is assumed to be $\textrm{PL}(\sqrt{d_k^2 + h^2}) \,{=}\, 1 {+} \left(\sqrt{d_k^2 + h^2}\right)^{\gamma}$ with $\gamma\,{=}\,2$ \cite{Ding17PoorRandBeamforming,Nadisanka_SPAWC}, and the UAV-BS altitude is $h\,{\in}\,[10,150]$~m.

\subsection{Location of the Most Detrimental Eavesdropper} \label{sec:Numerical-Location_Dist}

\begin{figure}[!t]
\vspace{-0.2in}
\centering
\includegraphics[width=0.45\textwidth]{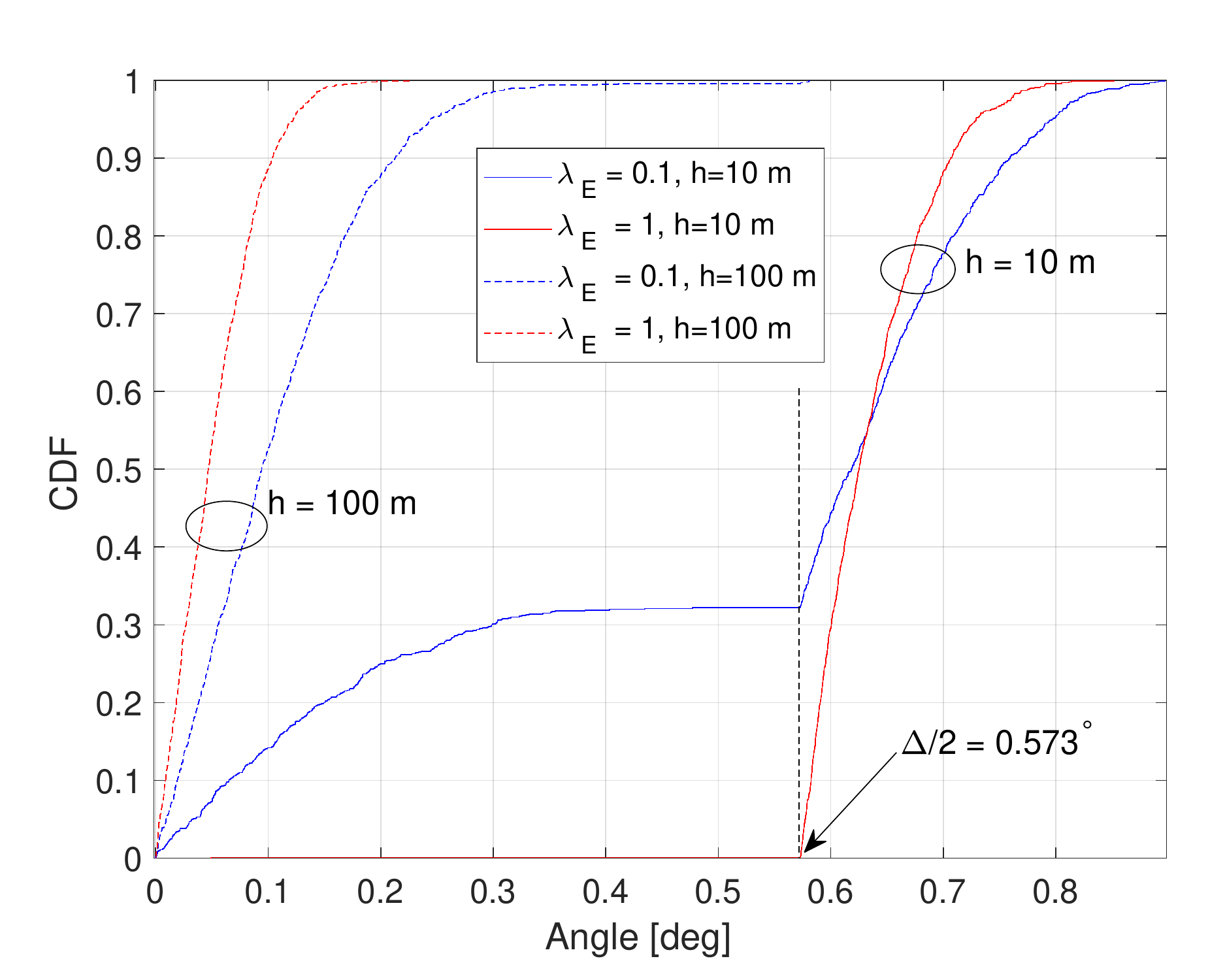}\vspace{0mm}
\caption{Angle distribution of the most detrimental eavesdropper.}
\label{fig:Angle_dist_most_det_Eve}
\end{figure}

\begin{figure}[!t]
\vspace{-0.1in}
\centering
\includegraphics[width=0.45\textwidth]{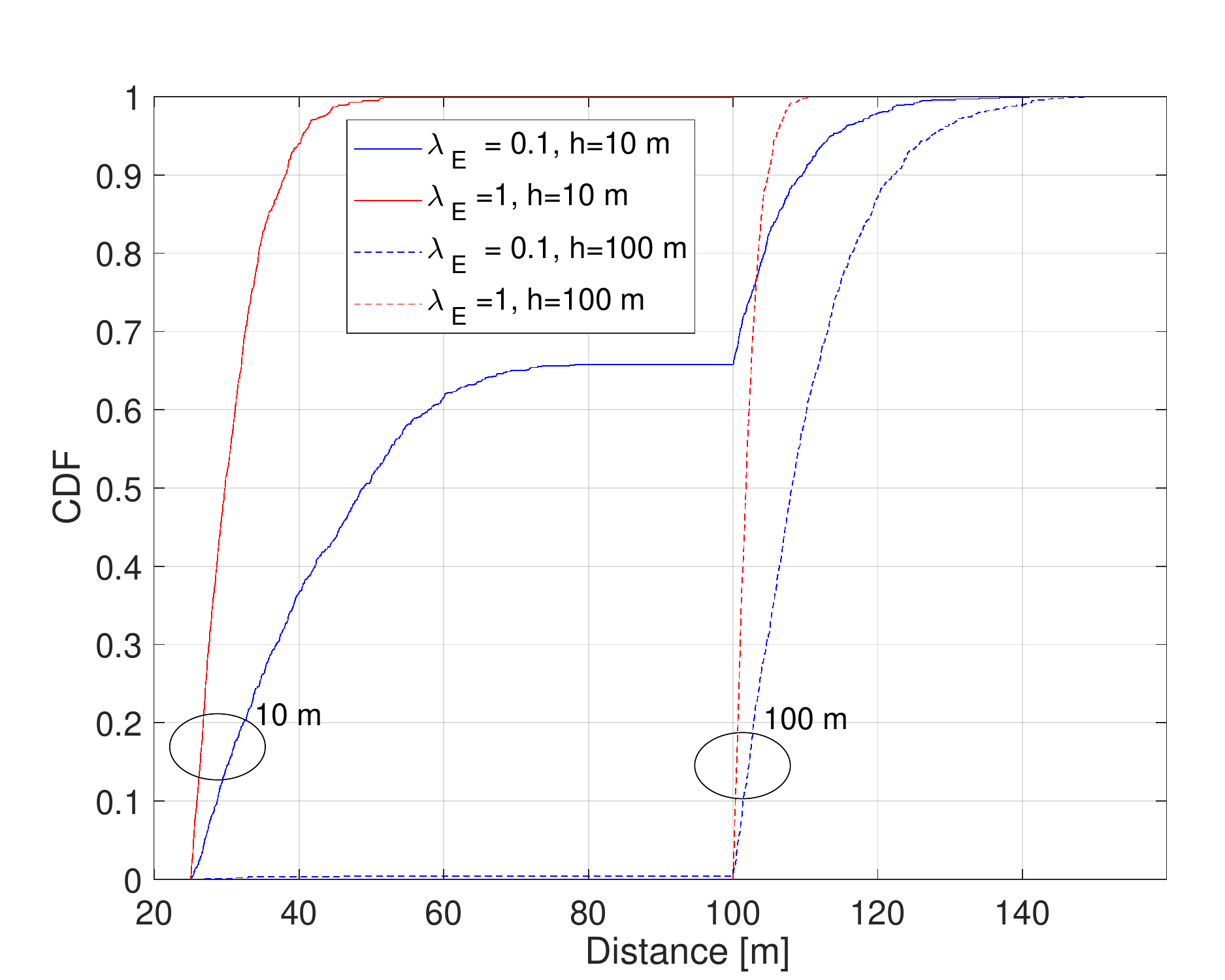}\vspace{0mm}
\caption{Distance distribution of the most detrimental eavesdropper.}
\label{fig:Distance_dist_most_det_Eve}
\end{figure}

We present the angle and distance distributions of the location of the most detrimental Eve in Fig.~\ref{fig:Angle_dist_most_det_Eve} and Fig.~\ref{fig:Distance_dist_most_det_Eve}, respectively, for two different altitudes of $h\,{=}\, \{10,100\}\,\text{m}$, and HPPP densities of $\lambda_{\rm E}\,{=}\,\{0.1,1\}$. In Fig.~\ref{fig:Angle_dist_most_det_Eve}, we observe that the most detrimental Eve is very likely to have a relative angle which is greater than $\frac{\Delta}{2}$ at a lower altitude of $h \,{=}\, 10\,\text{m}$. In particular, relative angle of the most detrimental Eve exceeds $\frac{\Delta}{2}$ all the time for $\lambda_{\rm E}\,{=}\,1$ while it drops to approximately $70\%$ of the time for $\lambda_{\rm E}\,{=}\,0.1$.
When the altitude becomes higher (i.e., $h \,{=}\,100\,\text{m}$), the relative angle of most detrimental Eve becomes smaller than $\frac{\Delta}{2}$. In Fig.~\ref{fig:Distance_dist_most_det_Eve}, we observe that the PL distance of the most detrimental Eve is smaller (greater) than $100\,\text{m}$ at lower (higher) altitudes, i.e., $h \,{=}\, 10\,\text{m}$ ($h \,{=}\, 100\,\text{m}$). We therefore conclude that \textit{the most detrimental Eve tends to have larger relative angles and smaller PL distances at lower altitudes in comparison to those at higher altitudes}.

\subsection{Impact of the Protected Zone Shape on Secrecy Rates}\label{sec:Numerical-Shape_Secrecy_Rates}

\begin{figure}[!t]
\vspace{-0.1in}
\centering
\includegraphics[width=0.45\textwidth]{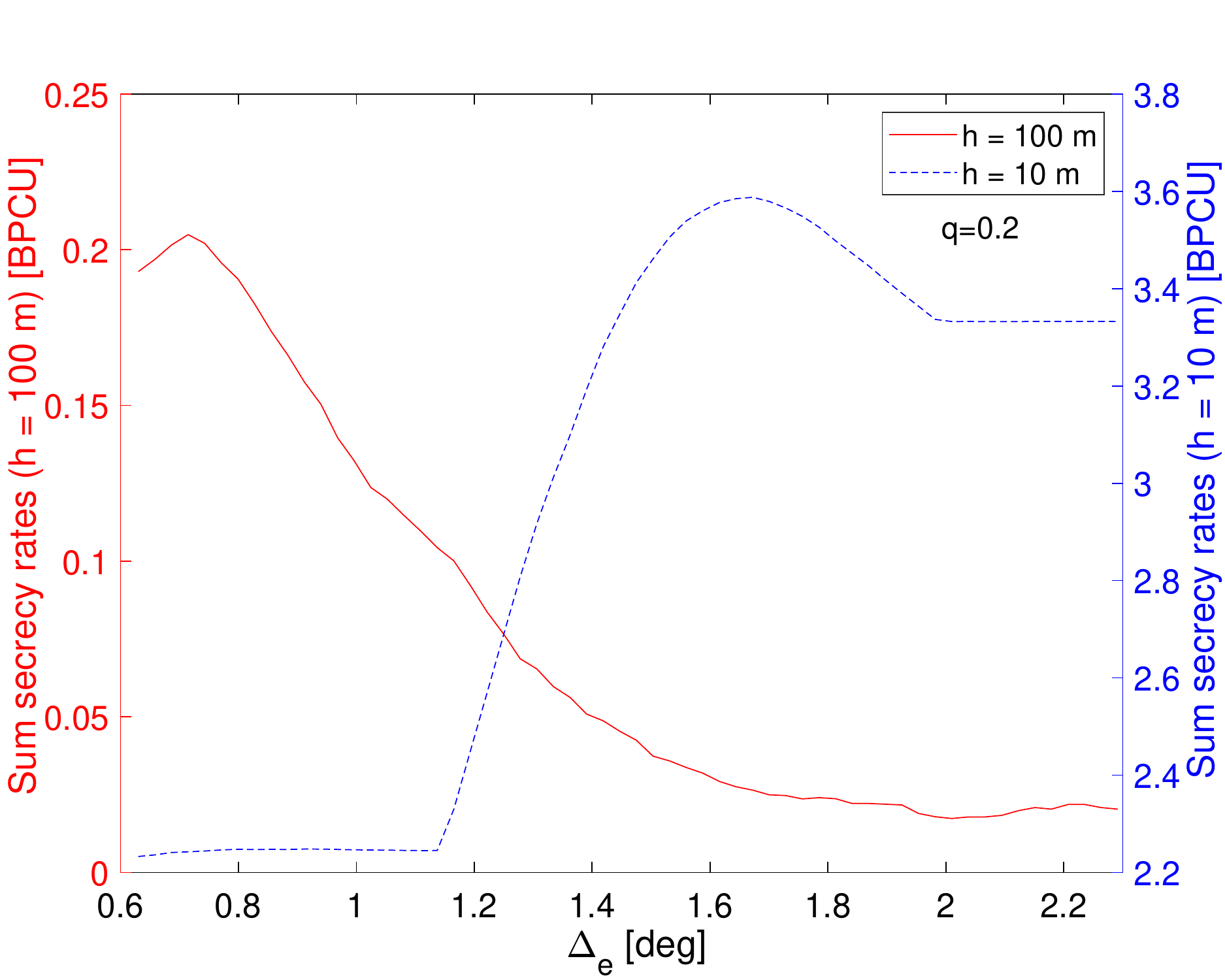}\vspace{0mm}
\caption{Sum secrecy rates of NOMA along with the protected zone angle (i.e., $\Delta_{\rm E}$) for $h\,{=}\, \{10,100\}\,\text{m}$, $q\,{=}\,0.2$, and $\lambda_{\rm E}\,{=}\,0.1$.}
\label{fig:Rate_Variation_Fixed_h_Delta_e}
\end{figure}

In Fig.~\ref{fig:Rate_Variation_Fixed_h_Delta_e}, we depict the sum secrecy rates along with the protected zone angle (i.e., $\Delta_{\rm E}$) at altitudes of $h \,{=}\, \{10,100\}\,\text{m}$ assuming $q \,{=}\, 0.2$. We observe that while the secrecy rates get maximized at $\Delta_{\rm E}\approx1.7^{\circ}\,(>\Delta)$ for $h \,{=}\, 10\,\text{m}$, the optimal angle turns out to be $\Delta_{\rm E}\approx 0.7^{\circ}\,(<\Delta)$ at $h \,{=}\, 100\,\text{m}$. This observation is consistent with the discussion in Section~\ref{sec:Numerical-Location_Dist} in the sense that the most detrimental Eve has a relative angle greater (smaller) than $\frac{\Delta}{2}$ at low (high) altitudes.

\begin{figure}[!t]
\vspace{-0.2in}
\centering
\includegraphics[width=0.45\textwidth]{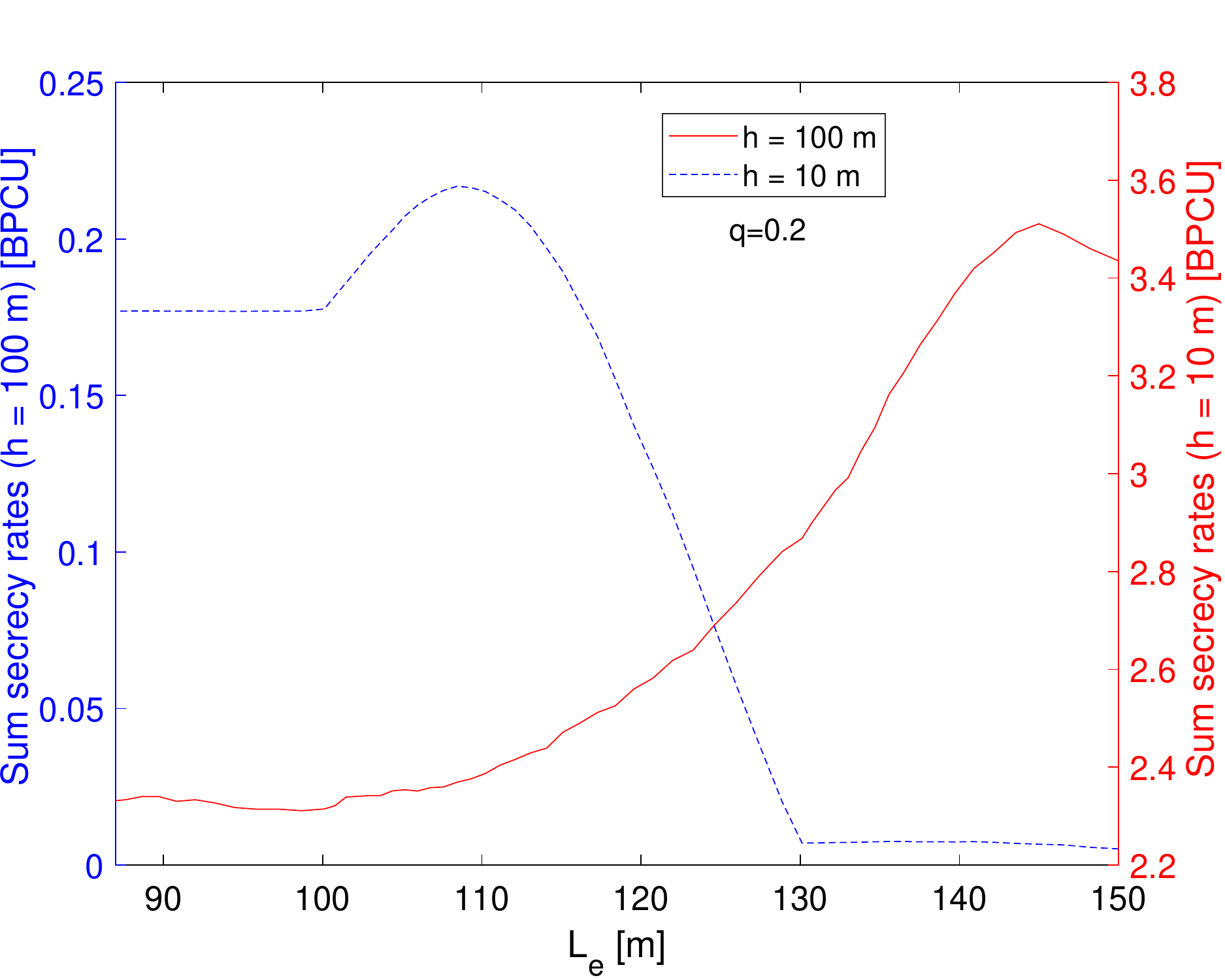}\vspace{0mm}
\caption{Sum secrecy rates of NOMA along with the protected zone distance (i.e., $L_{\rm E}$) for $h\,{=}\, \{10,100\}\,\text{m}$, $q\,{=}\,0.2$, and $\lambda_{\rm E}\,{=}\,0.1$.}
\label{fig:Rate_Variation_Fixed_h_L_e}
\end{figure}

Similarly, Fig.~\ref{fig:Rate_Variation_Fixed_h_L_e} presents the secrecy rates along with the protected zone distance (i.e., $L_{\rm E}$) for the same settings as of Fig.~\ref{fig:Rate_Variation_Fixed_h_Delta_e}. We observe that while the optimal distance maximizing the secrecy rates is $L_{\rm E} \,{\approx}\, 110 \, \text{m}$ at $h \,{=}\, 10\,\text{m}$, it turns out to be $L_{\rm E} \,{\approx}\, 145 \, \text{m}$ at $h \,{=}\, 100\,\text{m}$. As before, this observation also nicely agrees with our discussions  in Section~\ref{sec:Numerical-Location_Dist} regarding the distance distribution of the most detrimental Eve. This shows the importance of optimizing the protected zone shape at different hovering altitudes to maximize sum secrecy rates.

\subsection{Secrecy Rates Variation with Altitude} \label{sec:Numerical-Secrecy_Rates_vs_Altitude}


\begin{figure}[!t]
\vspace{-0.15in}
\includegraphics[width=0.45\textwidth]{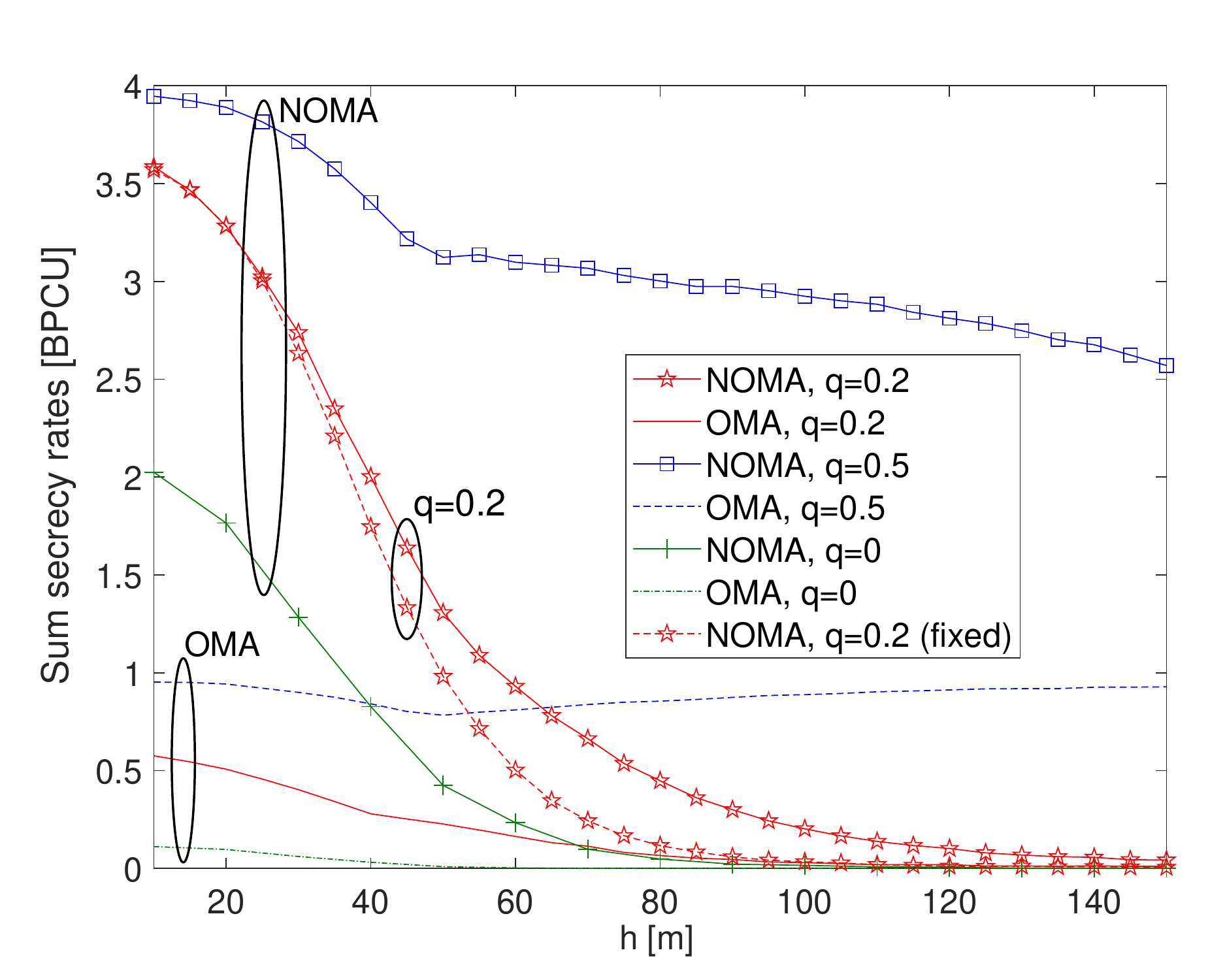}\vspace{0mm}
\caption{Sum secrecy rates for NOMA and OMA along with UAV-BS hovering altitude, where $q\,{\in}\, \{0,0.2,0.5\}$, and $\lambda_{\rm E}\,{=}\,0.1$.}
\label{fig:Rate_Variation_NOMA_OMA}
\end{figure}

In Fig.~\ref{fig:Rate_Variation_NOMA_OMA}, we present sum secrecy rates of NOMA and OMA transmission along with varying altitude of $h\,{\in}\,[10,150]\,\text{m}$ and for different protected zone sizes (i.e., $q\,{\in}\, \{0,0.2,0.5\}$). For a nonzero protected zone (i.e., $q\,{\neq}\, 0$), considering shape optimization as discussed in Section~\ref{sec:Shape_Optimization}, sum secrecy rates are identified. In addition, Fig.~\ref{fig:Rate_Variation_NOMA_OMA} also captures sum secrecy rate variation with $q\,{=}\, 0.2$ for a fixed shape (optimal shape at $h\,{=}\, 10$~m). As can be observed, the fixed protected zone shape yields sum secrecy rates comparable to that of optimized protected zone shape only around $h\,{=}\, 10$~m and performs worse at all the other altitudes. Further, we observe that the secrecy rates improve if large portion of the Eve region can be covered by the protected zone (i.e., $q$ increases). Based on the target sum secrecy rate and the operational altitude, the smallest $q$ can also be determined. By this way, the desired secrecy rates can be achieved optimally by designating less area as the protected zone which would relieve the burden of clearing any unnecessary region free from Eves. Note also that the secrecy rates associated with NOMA is much larger than those of OMA especially at lower altitudes.

\begin{figure}[!h]
\vspace{-0.2in}
\centering
\subfloat[Optimal angle, $\Delta_{\rm E}^{*}$.]{\includegraphics[width=0.25\textwidth]{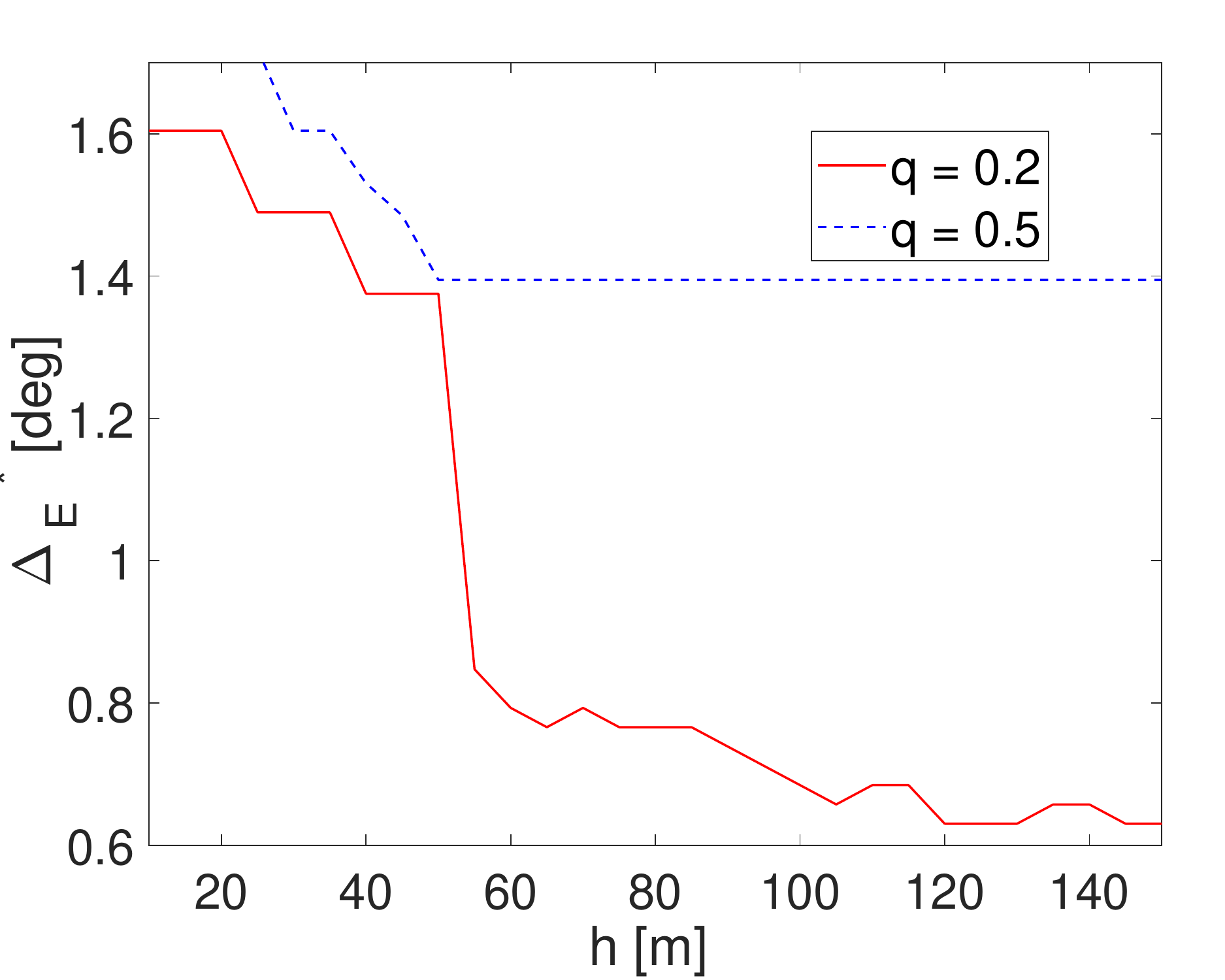}
\label{fig:anlge_dist}}
\subfloat[Optimal distance, $L_{\rm E}^{*}$.]{\includegraphics[width=0.25\textwidth]{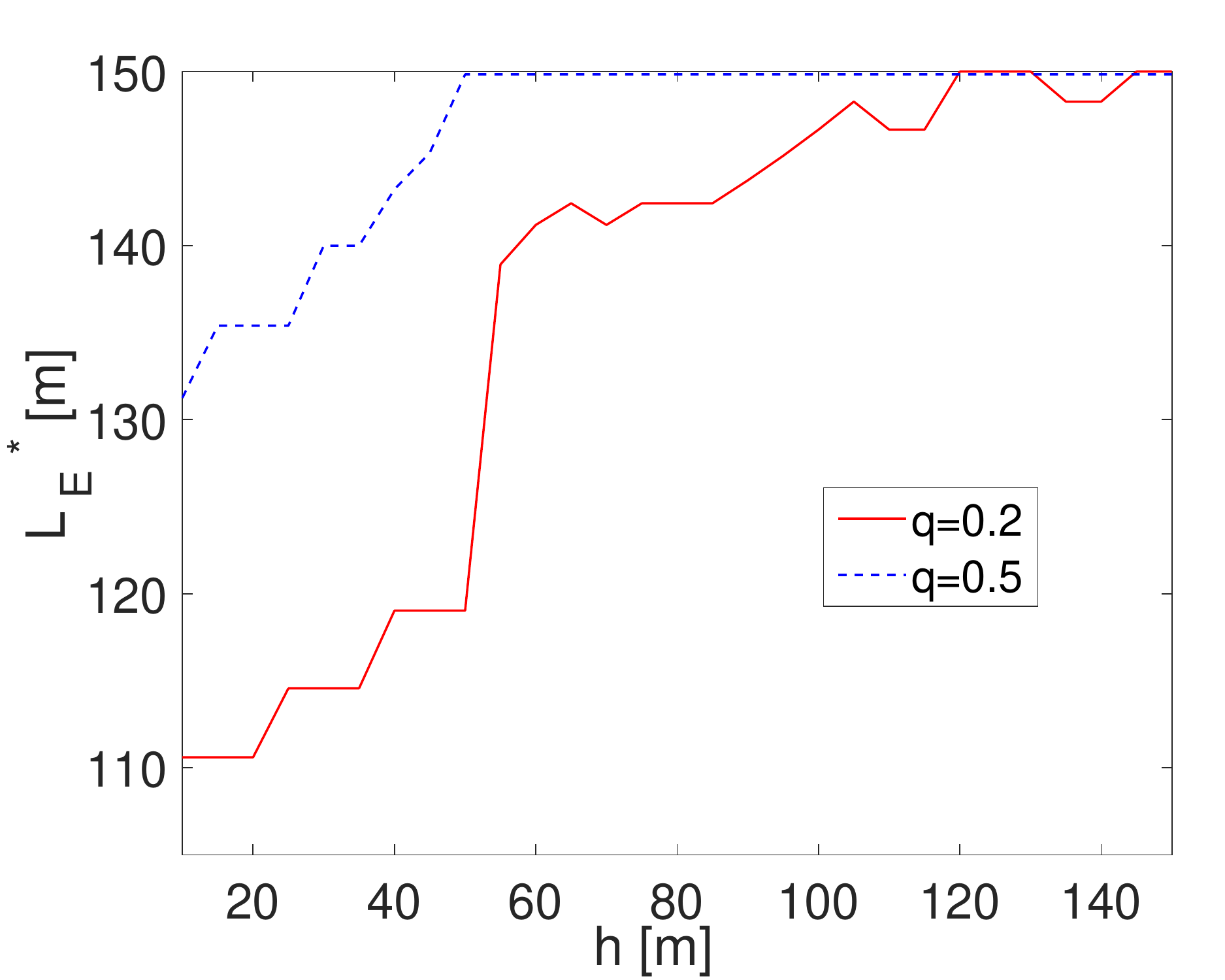}
\label{fig:distance_dist}}
\vspace{0.1in}
\caption{$\Delta_{\rm E}^{*}$ and $L_{\rm E}^{*}$ variation with varying UAV-BS hovering altitudes, Here $q\,{=}\,0.2, 0.5$.}
\label{fig:Optimal_angle_distance_variation}
\end{figure}

In Fig.~\ref{fig:Optimal_angle_distance_variation}, variation of the optimal shape of the protected zone is captured for $q\,{=}\,0.2, 0.5$. In particular, Fig.~\ref{fig:anlge_dist} shows the optimal angle, $\Delta_{\rm E}^{*}$ variation whereas Fig.~\ref{fig:distance_dist} depicts optimal distance $L_{\rm E}^{*}$ variation with UAV-BS hovering altitude. As can be observed from Fig.~\ref{fig:Optimal_angle_distance_variation}, $\Delta_{\rm E}^{*}$ decreases with altitude (see Fig.~\ref{fig:anlge_dist}) while $L_{\rm E}^{*}$ increases with altitude (see Fig.~\ref{fig:distance_dist}). This observation aligns nicely with the discussion in Section~\ref{sec:Numerical-Location_Dist} which tells us that at lower altitudes the most detrimental Eve tends to have a larger relative angle and smaller distance whereas at higher altitudes this is vice versa.

\section{Concluding Remarks}\label{sec:conclusion}
In this paper, we investigate the secrecy rates of UAV based mmWave communication network considering NOMA transmission. In particular, we consider \textit{protected zone} approach to enhance the secrecy rates. Towards this end, we first investigate the distribution of the location of the most detrimental Eve which impairs secrecy rates the most. We then consider the protected zone which is free from any Eve, and the associated optimal shape of it to enhance the secrecy performance. We show that the optimal shape of the protected zone should cover the most detrimental Eve. In addition, we also show that the optimal shape highly relies on the UAV-BS hovering altitude such that the protected zone should be wider (narrower) in angle and shorter (longer) in distance at lower (higher) altitudes.


\bibliographystyle{IEEEtran}
\bibliography{Doc_Ref}
\end{document}